# Distributed vibration sensing based on forward transmission and coherent detection

YAXI YAN,[3,1] CHANGJIAN GUO,[1,3,*] XIONG WU,[3] ZIQI LIN,[1] XIAN ZHOU,[2] FAISAL NADEEM KHAN,[4] ALAN PAK TAO LAU,[4] AND CHAO LU[3]

[1]*South China Academy of Advanced Optoelectronics, South China Normal University, Guangzhou, China*
[2]*Research Center for Convergence Networks and Ubiquitous Services, University of Science & Technology Beijing (USTB), No.30 Xue Yuan Road, Haidian, Beijing, 100083, China*
[3]*Department of Electronic and Information Engineering, The Hong Kong Polytechnic University, Hung Hom, Kowloon, Hong Kong (SAR), China*
[4]*Department of Electrical Engineering, The Hong Kong Polytechnic University, Hung Hom, Kowloon, Hong Kong (SAR), China*
*\*changjian.guo@coer-scnu.org*

**Abstract:** A novel ultra-long distributed vibration sensing (DVS) system using forward transmission and coherent detection is proposed and experimentally demonstrated. In the proposed scheme, a pair of multi-span optical fibers are deployed for sensing, and a loop-back configuration is used by connecting the two fibers at the far end. The homodyne coherent detection is used to retrieve the phase and state-of-polarization (SOP) fluctuations caused by a vibration while the localization of the vibration is realized by tracking the phase changes along the two fibers. The proposed scheme has the advantage of high signal-to-noise ratio (SNR) and ultra-long sensing range due to the nature of forward transmission and coherent detection. In addition, using forward rather than backward scattering allows detection of high frequency vibration signal over a long sensing range. More than 50dB sensing SNR can be obtained after long-haul transmission. Meanwhile, localization of 400 Hz, 1 kHz and 10 kHz vibrations has been experimentally demonstrated with a spatial resolution of less than 50 m over a total of 1008 km sensing fiber. The sensing length can be further extended to even trans-oceanic distances using more fiber spans and erbium-doped fiber amplifiers (EDFAs), making it a promising candidate for proactive fault detection and localization in long-haul and ultra-long-haul fiber links.

## 1. Introduction

Driven by the bandwidth-consuming applications such as video streaming, cloud computing, and proliferation of smart devices, the data traffic in long-haul, metro and optical access networks has increased dramatically. The survivability has increasingly become a challenge for these optical networks due to the huge internet traffic. It is therefore imperative for network operators to have an early warning and proactive protection mechanism incorporated into their networks. Traditional network monitoring uses optical time-domain reflectometry (OTDR) schemes which can only localize the fault event without knowledge of the root cause of the failure [1-2]. Recently, fault detection and classification schemes by monitoring the state of polarization (SOP) rotation using either a commercial polarimeter [3] or a standard coherent receiver [4] were proposed. However, such schemes still lack the capability to localize the fault events.

Fiber-optic vibration sensors can be used for leakage detection of oil and gas pipelines, structure health monitoring, perimeter security protection, and fault detection in optical communication links. In the past few decades, distributed vibration sensors (DVS) have attracted much research attention because of their advantages like resistance against electromagnetic interference, high precision, good chemical stability, and long sensing distance [5]. The DVS can be mainly divided into two categories. First is based on the OTDR technique, including phase-sensitive OTDR [6,7], polarization OTDR [8] and Brillion scattering based

OTDR [9]. Another is based on the interferometer technique, including Michelson interferometer (MI) [10], Mach-Zehnder interferometer (MZI) [11,12] and Sagnac interferometer [13]. In OTDR based DVS systems, since the back-scatterred light is very small, the sensing distance and spatial resolution are limited. In [14], a hybrid amplification scheme was proposed to extend the sensing distance of OTDR-based DVS and consequently a 175km sensing range was obtained with 25m spatial resolution. Meanwhile, since the frequency response of vibration is related to the pulse repetition rate, the dynamic response will be a problem in long distance OTDR-based DVS. In interferometric DVS systems, the detectable frequency is no more a problem. However, most of these systems are not appropriate for long sensing distance because of the influence of the Rayleigh backscattering noise. For Sagnac interferometric DVS, a 41km sensing range with 100m spatial resolution was reported [15]. For MI-based DVS, a 51m spatial resolution was achieved over 4012m of sensing fiber [10]. In [16], a record of 320km sensing distance using MZI was reported. It should be noted that MZI-based DVS requires a stable reference arm as well as polarization tracking.

In this paper, we propose and experimentally demonstrate a novel DVS system using forward transmission and homodyne coherent detection. A pair of multi-span optical fibers are deployed at the same location and utilized for sensing while a loop-back configuration is used by connecting the two fibers at the far end. Since both fibers are placed at the same location, the same phase change patterns will occur in case of disturbance events. Localization of these events are done by determine the time delay between the two patterns. Compared with the DVS systems mentioned above, the proposed structure is much simpler. A commercial phase- and polarization-diversity coherent receiver is used to extract the phase and SOP information from the received signal after forward transmission, which is then used for detection and localization of the vibration incident. An optical fiber link with a total length of 1008 km is investigated in this work. The obtained spatial resolution is less than 50 meters. The sensing length can be further extended to several thousand kms by using even more spans of fibers and EDFAs. Therefore, the proposed scheme may have applications in long-haul and ultra-long-haul optical fiber communication links, as long as the fiber links in both directions share the same optical cables.

## 2. Operating principle

### 2.1 Coherent detection

In our experiment, a commercial phase- and polarization-diversity coherent receiver is used for detection and localization of the received signals after forward transmission through the fiber under test (FUT). The four-channel output signals of the receiver contain in-phase ($I_x$, $I_y$) and quadrature ($Q_x$, $Q_y$) components of the homodyne beating signal between the forward transmitted light and the local oscillator (LO), in both $x$- and $y$- polarizations, which can be expressed as the following:

$$I_x = R\sqrt{\frac{\alpha P_s P_{LO}}{2}} \cos(\theta_x(t) - \theta_{LO}(t) + \delta) \tag{1}$$

$$Q_x = R\sqrt{\frac{\alpha P_s P_{LO}}{2}} \sin(\theta_x(t) - \theta_{LO}(t) + \delta) \tag{2}$$

$$I_y = R\sqrt{\frac{(1-\alpha) P_s P_{LO}}{2}} \cos\left(\theta_y(t) - \theta_{LO}(t)\right) \tag{3}$$

$$Q_y = R\sqrt{\frac{(1-\alpha) P_s P_{LO}}{2}} \sin\left(\theta_y(t) - \theta_{LO}(t)\right) \tag{4}$$

where $P_s$, $P_{LO}$ are the powers of the forward transmitted light and $R$ is the responsivity of the photodetectors in the coherent receiver, $\alpha$ is the power ratio of the two polarization components, and $\delta$ is the phase difference between them. $\theta_x(t)$, $\theta_y(t)$ and $\theta_{LO}(t)$ are the phases of the forward transmitted x-polarization, y-polarization signals and LO, respectively. From Eqs. (1)-(4), one can calculate the phase difference $\Delta\theta(t) = \theta_x(t) - \theta_{LO}(t)$ between the forward transmitted light and LO. Since the phase change is directly detected and obtained, we cannot only realize the event detection (e.g., intrusion detection) but also quantify the event, such as measuring the exact frequency and amplitude of the external environmental perturbations.

### 2.2 Localization principle

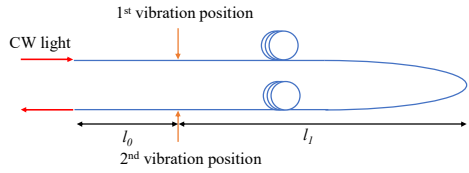

Fig. 1. Schematic diagram of the location principle.

When light passes through an FUT, the phase of the output light $\varphi$ can be expressed as $\varphi = \frac{2\pi n L}{\lambda_0} + \varphi_0$, where $\lambda_0$ is the wavelength of light in vacuum, $n$ is the refractive index of the fiber core, $L$ is the length of FUT, $\varphi_0$ is the original phase of the CW light output from the laser. As shown in Fig. 1, the FUT consists of two fibers placed together along the sensing area. These two fibers are put close to each other so that the same phase shift and SOP rotation would happen for both fibers in case of vibration events. The far ends of the two fibers are spliced together to form a loop-back arrangement. When there is no external vibration, the phase received at the output point will only slowly change because of the intrinsic phase noise of the laser. At time $t_0$, an external vibration signal is applied on the FUT at a certain position. The refractive index of the fiber core and the fiber birefringence will change consequently and hence the phase of the propagating light will be largely changed at the vibration position. Due to the loop-back configuration, a single external vibration will affect two different positions of the FUT. At time $t_0 + nl_0/c$, where $c$ is the velocity of light in vacuum, there will appear a sharp phase change in the received signal, which is caused by the vibration at 2nd position. At time $t_0 + nl_0/c + 2nl_1/c$, there will appear a second sharp phase change in the received signal, which is caused by the vibration at 1st position. The time delay $T$ between the first and second phase change is $2nl_1/c$. By measuring the time delay $T$, the fiber length $l_1$ can be obtained. As the whole length of FUT is already known, the vibration position can be located. Meanwhile, the amplitude of the vibration can be obtained by retrieving the vibration-induced phase shift after a calibration is performed.

### 3. Experimental setup

Figure 2(a) shows the experimental setup of the long-range DVS system. The output of an ultra-narrow-linewidth CW tunable laser operating at 1550 nm is split into two branches. The upper branch is launched into the FUT directly. The total length of the FUT is around 1008 km, including two 100-km spans and ten 80-km spans. At the end of each fiber span, an EDFA and a 100 GHz band pass filter (BPF) are used to compensate the link loss and remove the out-of-band noise, respectively. At both the beginning and the end of the FUT, two 10m fibers are wrapped around the same PZT to cause the external vibrations. The CW signal in the lower branch is sent to the CR serving as the LO. A polarization controller is used to adjust the polarization state of the forward transmitted light before the CR. Here, we use a phase- & polarization-diversity CR to receive the signal. The four-channel electrical signals are then sampled by a real-time oscilloscope with a sampling rate of 2 Gsamples/s. The offline digital signal processing (DSP) is then used to analyze the collected data. It should be noted that the SOP changes could also be tracked and restored in our setup since a polarization diversity CR is utilized. Furthermore, the tracked SOP information could be used for detection and localization of the vibration events. However, for simplicity and without loss of generality, a polarization controller (PC) is used in this experiment to align the SOP of the signal to that of the LO, since the vibration-induced phase changes are much more significant compared with the SOP variations in case of vibrations [17].

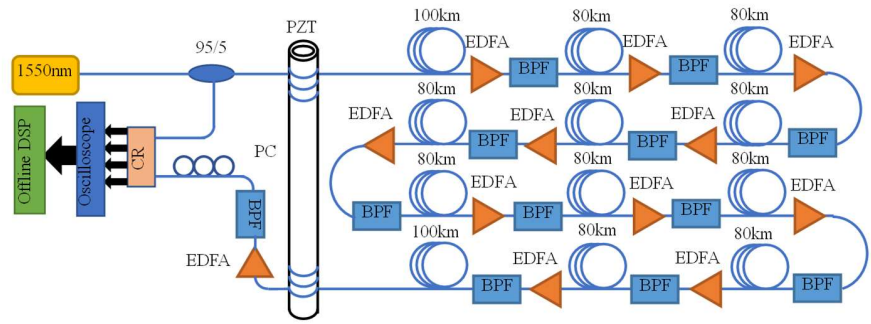

(a)

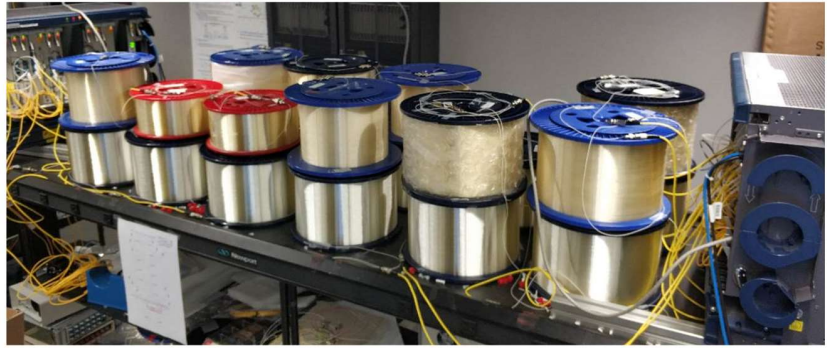

(b)

Fig. 2. (a) The configuration of the proposed system; (b) picture of the 1008 km fiber link. PZT: piezoelectric ceramic transducer, BPF: band pass filter, CR: coherent receiver.

### 4. Experimental results

Firstly, the localization performance under vibration events of different frequencies is investigated. Sinusoidal waves with three different frequencies, i.e., 400 Hz, 1 kHz and 10 kHz, are used to drive the PZT using an applied peak-to-peak voltage of 10 V. If there is no vibration, the received signal after PD should be a slow-varying current. When an external vibration is applied, abrupt changes in the amplitudes of the received signals may happen due to the vibration-induced transient phase changes. The oscilloscope will then be triggered to collect the fast-varying received signals. The received complex signals are then low pass-filtered and their optical phases are extracted.

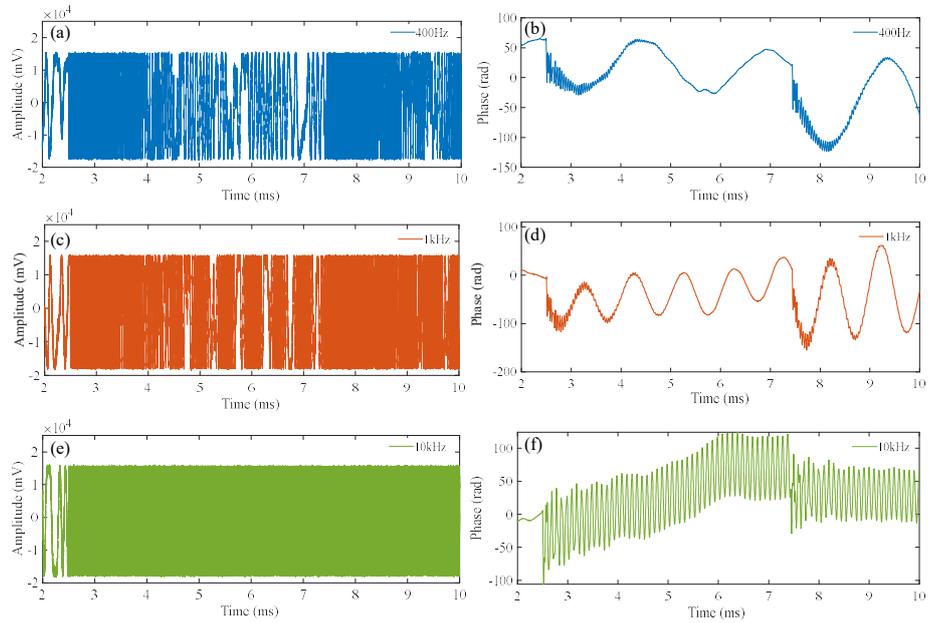

Fig. 3. (a), (c), (e) Amplitudes of signals with frequencies of 400 Hz, 1 kHz and 10 kHz during the 8 ms period. (b), (d), (f) Phase information demodulated from the signals with frequencies of 400 Hz, 1 kHz and 10 kHz.

Figures 3. (a), (c) and (e) show the amplitude curves of the detected signals with frequencies of 400 Hz, 1 kHz and 10 kHz, respectively. The demodulated phases of the detected signals are shown in Figs. 3(b), (d) and (f). The following observations can be made. a) It is obvious that the amplitude of the signal changes much faster after the vibration occurs. b) There are two abrupt phase shifts on each of the phase curves, which are induced by the vibrations along the FUT. c) High frequency components on the phase curves can be found right after the abrupt phase changes due to the PZT-induced strain on the fibers [18,19]. d) Gradually, the phase presents a clear sinusoidal variation which matches well with the sinusoidal signal applied on the PZT.

In Fig. 4, the phase gradient curves for three different frequencies and their corresponding cross correlation curves are presented. Here, the sampling rate is 2 Gsamples/s, so the corresponding time resolution is 0.5 ns. It can be seen that compared to the phase curves shown in Figs. 3(b), (d) and (f), the phase gradient curves reflect the vibration patterns much more clearly. Two distinct vibration patterns can be seen in each of the phase gradient curves. Since the two vibrations are generated by the same PZT, the variation patterns of the phase gradient curves are similar. Therefore, by calculating the cross correlation of the two vibration-induced phase gradient patterns, one can easily obtain the vibration position by calculating the time delay between the two peaks. From Fig. 4(b), the time stamps of the two peaks are 5.0000000 ms and 9.9368140 ms, respectively, with a time delay of 4.9368140 ms, indicating that the distance between the two vibration positions is 1008187.7 m. Similarly, from Fig.4(d) and (f), the calculated length between the two vibration positions is 1008127.8 m and 1008166.2 m, respectively. We run the localization test 8 times for each frequency and the results are shown in Table 1. One can see from Table 1 that the standard deviation of the localization results is smaller for larger vibration frequencies, and that the spatial resolution is less than 50 m for a total sensing length of around 500 km.

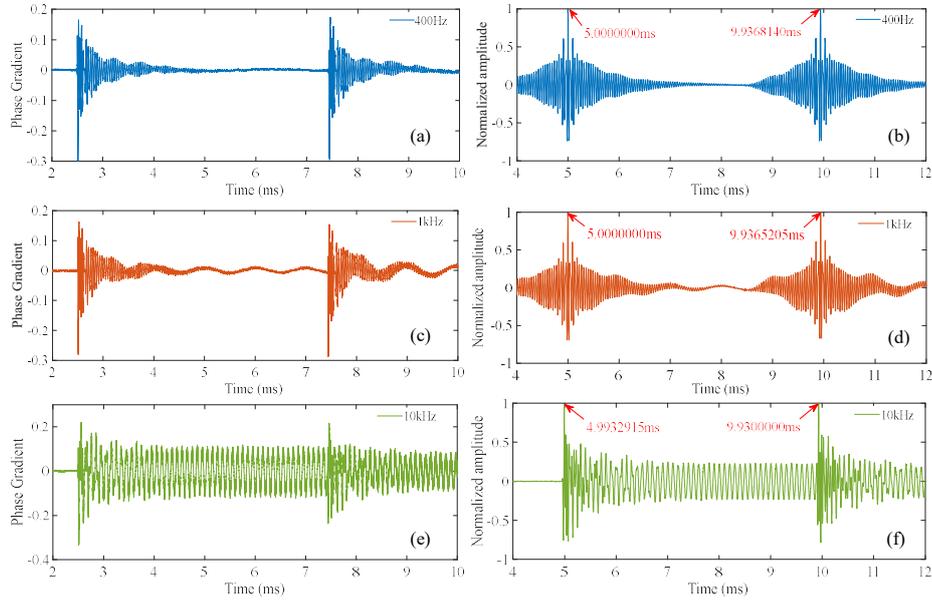

Fig. 4. (a), (c), (e) Phase gradient curves of the signals with frequencies of 400 Hz, 1 kHz and 10 kHz. (b), (d), (f) Cross correlation curves obtained from the corresponding phase gradient curves.

Table 1. Experimental results of the vibration localization.

| Vibration frequency (Hz) | | 400 | | 1 k | | 10 k |
|---|---|---|---|---|---|---|
| Distance (m) | 1 | 1008187.7 | 1 | 1008127.8 | 1 | 1008166.2 |
| | 2 | 1008169.3 | 2 | 1008149.1 | 2 | 1008178.2 |
| | 3 | 1008204.2 | 3 | 1008166.3 | 3 | 1008182.4 |
| | 4 | 1008136.7 | 4 | 1008203.8 | 4 | 1008159.4 |
| | 5 | 1008223.9 | 5 | 1008172.6 | 5 | 1008150.2 |
| | 6 | 1008210.4 | 6 | 1008144.0 | 6 | 1008175.5 |
| | 7 | 1008158.6 | 7 | 1008145.8 | 7 | 1008170.6 |
| | 8 | 1008207.6 | 8 | 1008212.2 | 8 | 1008171.6 |
| Standard deviation (m) | | 28.01579911 | | 27.92557072 | | 9.802160157 |

We also measured the vibration-induced phase variations as a function of the driving voltage, as shown in Fig. 5(a). The vibration frequency is fixed to 10 kHz. We changed the voltage applied to the PZT from 1 V to 10 V. The FUT is exposed to the external environment so that the environmental influence, e.g., wind and temperature variations will cause a slow change in the phase, resulting in a slow varying envelope of the 10 kHz sinusoidal waveform. The relationship between the amplitude of the phase variation and the voltage applied to the PZT is shown in Fig. 5(b). The value of $R^2$ is 0.9996.

To prove the broad frequency response of this sensing system, we also investigated its frequency response by applying several different vibration frequencies with a peak-to-peak voltage of 10 V. The results are shown in Fig. 6(a). The measured SNRs for 200 Hz, 600 Hz,

1kHz, 5 kHz and 10 kHz vibrations are 26 dB, 39 dB, 48 dB, 62 dB and 78 dB, respectively. It is found that the SNR decreases with the frequency due to the residual carrier phase noise. Although homodyne coherent detection is utilized in the experiment, the phase noise still exists due to the ultra-long sensing fibers. If the length of the sensing fiber is within the coherence length of the laser, which is the case in our experiment since the coherence length of our 100Hz fiber laser is $L_c = c/(n_{fiber} \Delta f) > 2000$ km, the carrier phase noise after coherent detection is subject to a normal distribution with a standard deviation $\sigma \propto \Delta f \times L$. The measured minimum frequency in our experiment is around 140 Hz, as shown in Fig. 6 (b). One can see that when the frequency is further decreased to 100 Hz, the phase signal cannot be identified.

The experimental results shown in Figs. 4, 5 and 6 clearly shows that ultra-long-range vibration sensing can be achieved with a very good SNR using the proposed scheme. The proposed scheme utilizes forward transmission which is fully compatible with existing optical communication networks. Due to hardware constraints, the demonstrated sensing distance is around 1008 km in this work. However, it is obvious that the sensing range of the proposed scheme can be extended to trans-oceanic distances, making it ideal for long-haul and ultra-long-haul applications.

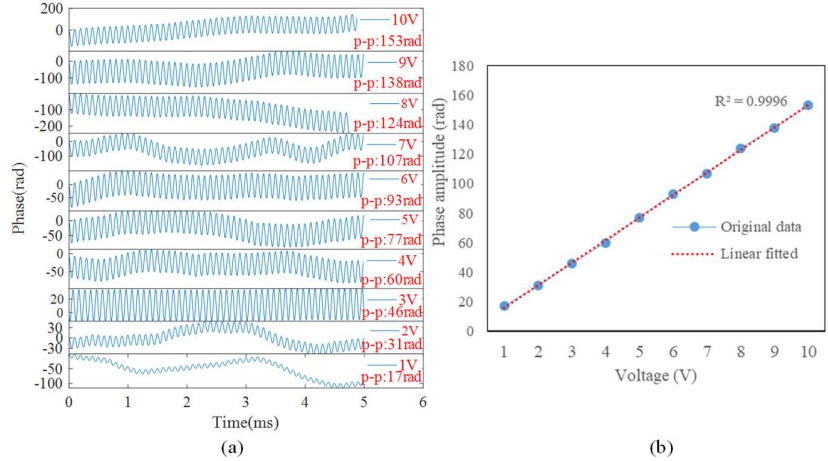

Fig. 5. (a) Stable sinusoidal vibration signal. The voltage applied to the PZT is tuned from 1 V to 10 V. (b) The relationship between the amplitude of phase variation and the voltage applied to the PZT.

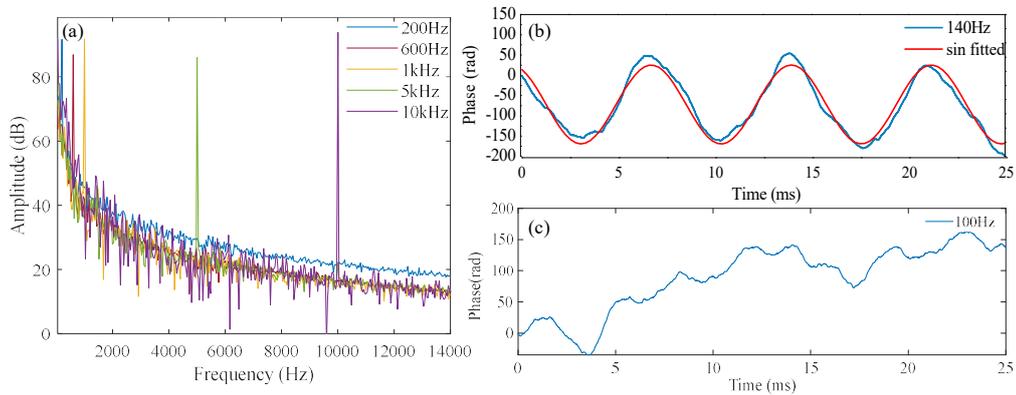

Fig. 6. (a) Spectra for different vibration frequencies. Phase information demodulated from the signals with frequency of (b) 140 Hz and (c) 100 Hz.

## 5. Conclusions

We have demonstrated a novel optical vibration sensing technique based on forward transmission and coherent detection with record long sensing range, using a pair of optical fibers deployed at the same location with loop-back configuration. Detection and localization of 400 Hz, 1 kHz and 10 kHz vibration signals have been successfully demonstrated in a 1008 km multi-span optical fiber channel, with a spatial resolution of less than 50 m and a measured SNR of more than 50 dB. A good linearity of the vibrations-induced phase changes has been demonstrated. Compared with the previously reported DVS schemes, the proposed structure is much simpler and also demonstrates ultra-long sensing range, thereby making the proposed method a promising candidate for proactive fault detection and localization in long-haul optical fiber links.

## Funding

National Natural Science Foundation of China (NSFC) (61435006, 61671053, 61871030); Natural Science Foundation (NSF) of Guangdong Province (2018A0303130117); Pearl River Science and Technology Nova Program of Guangzhou (201710010028); The Research Grants Council (RGC) of Hong Kong, General Research Fund (GRF: PolyU 152658/16E, 152168/17E); and The Hong Kong Ph.D. Fellowship.